# The role of potential energy landscape research in the development of new electrolyte solutions

Vitaly V. Chaban

Yerevan State University, Yerevan, 0025, Armenia. E-mail: vvchaban@gmail.com

**Abstract.** The development of new electrolyte solutions with improved characteristics is a key challenge for creating high-performance batteries, fuel cells, supercapacitors, and other electrochemical devices. The study of the potential energy landscape (PEL) plays an important role in this process, providing information about the interactions between solution components at the molecular level. In this work, we review the practice of applying PEL research methods based on classical and quantum-chemical algorithms to analyze the structure, dynamics, and thermodynamic properties of electrolyte solutions. Intermolecular and ion-molecular interactions at the microscopic level, which determine the macroscopic properties of the electrolyte solution, are considered in detail. The importance of identifying stable configurations of ions and their solvates is emphasized. PEL analysis allows for the systematic determination of the most probable structures and complexes formed in solution, which is important for understanding ion transport mechanisms. The study of the PEL allows for the determination of the energy barriers that must be overcome for ion migration, which is related to the conductivity of the electrolyte. The application of PEL research methods in combination with experimental data opens up new possibilities for the rational design of electrolyte solutions with desired physicochemical properties.

# Роль исследования ландшафта потенциальной энергии в разработке новых электролитных растворов

Виталий Витальевич Чабан

Ереванский государственный университет, Ереван 0025, Армения. E-mail: vvchaban@gmail.com.

**Аннотация**. Разработка новых электролитных растворов с улучшенными характеристиками является ключевой задачей при создании высокоэффективных аккумуляторов, топливных элементов, суперконденсаторов и других электрохимических устройств. Исследование ландшафта потенциальной энергии (ЛПЭ) играет важную роль в этом процессе, предоставляя информацию о взаимодействиях между компонентами раствора на молекулярном уровне. В данной работе мы рассматриваем практику применения методов исследования ЛПЭ, основанных на классических и квантово-химических алгоритмах, для анализа структуры, динамики и термодинамических свойств электролитных растворов. Подробно рассматриваются межмолекулярные и ион-молекулярные взаимодействия на микроскопическом уровне, определяющие макроскопические свойства электролитного раствора. Подчеркнута важность идентификации стабильных конфигураций ионов и их сольватов. Анализ ЛПЭ позволяет систематически определять наиболее вероятные структуры и комплексы, образующиеся в растворе, что важно для понимания механизмов ионного транспорта. Исследование ЛПЭ позволяет определить энергетические барьеры, которые необходимо преодолеть для миграции ионов, что связано с проводимостью электролита. Применение методов исследования ЛПЭ в сочетании с экспериментальными данными открывает новые возможности для рационального дизайна электролитных растворов с требуемыми физико-химическими свойствами.

**Ключевые слова**: электролитный раствор; молекулярное моделирование; молекулярная динамика; квантовая химия; ландшафт потенциальной энергии.

# Введение

Из-за стремительного роста количества бытовых устройств, питающихся от электросети, современное человечество все более и более активно нуждается в дополнительных и более эффективных источниках энергии. Это замечание в равной степени касается как первичных источников энергии, так и перезаряжаемых устройств, среди которых отдельного упоминания заслуживает портативная электроника (телефоны, планшеты, ноутбуки) [1-4].

Концепция ландшафта потенциальной энергии (ЛПЭ) представляет собой воображаемую многомерную поверхность, которая описывает потенциальную энергию системы в зависимости от внутренних координат ее составляющих (рис. 1). В контексте электролитных растворов ЛПЭ отражает взаимодействие между ионами, молекулами растворителя, электродам, технологически обусловленными добавками и другими компонентами системы. Систематический анализ ЛПЭ позволяет получить ценную информацию о возможных микроскопических состояниях раствора, механизмах сольватации и транспорта ионов в ходе зарядки и разрядки, а также термодинамических свойствах [5-6]. В качестве энергии ЛПЭ может использовать потенциальную энергию системы, либо полную энергию системы, либо иные термодинамические потенциалы в зависимости от искомого варианта интерпретации концепции. Использование полной энергии системы позволяет изучать электролитные системы в условиях конечных температуры и давления.

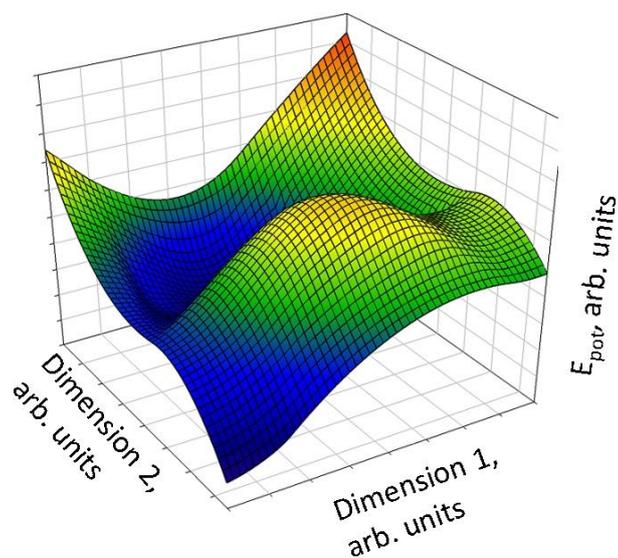

Рис. 1.Модельный ЛПЭ для малого участка произвольной электролитной системы.

В настоящее время наблюдается стремительный рост исследований в области электрохимических устройств, таких как литий-ионные аккумуляторы [7-9], натрий-ионные аккумуляторы [10-11], магний-ионные аккумуляторы [12], твердотельные аккумуляторы, топливные элементы [13-15], суперконденсаторы [16-18] и электролизеры [19-21]. Этот академический интерес обусловлен стабильно возрастающей потребностью в производстве эффективных и устойчивых системах хранения и преобразования энергии [22-25]. Помимо локальных нужд электрохимические устройства естественным образом играют важную роль при решении таких глобальных проблем цивилизации, как изменение климата и истощение запасов ископаемого топлива. Электролиты, выступая в качестве среды для переноса ионов между электродами, играют ключевую роль в эффективности функционирования и долговечности этих устройств. Электролиты напрямую влияют на важнейшие эксплуатационные характеристики электрохимических устройств, такие как эффективность, мощность, срок службы, рабочий диапазон температур и безопасность [26-27].

Отдельным требованием современного общества является экологичность или хотя бы относительная экологичность используемых аккумуляторов. Цена переработки химического источника тока должна быть включена в общую стоимость разрабатываемой в современном глобальном мире энергетической технологии. Наряду с экологичностью важное место занимает вопрос безопасности аккумулятора в контексте конечного потребителя продукции.

Существующие на данный момент электролиты зачастую не в полной мере удовлетворяют требованиям современных электрохимических устройств [27]. Например, широко используемые в литий-ионных аккумуляторах органические электролиты обладают ограниченной ионной проводимостью и пропорционально высокой вязкостью, узким электрохимическим окном, высокой воспламеняемостью и токсичностью. Высоковязкие электролитные растворы создают дополнительные сложности при их разработке и производстве. Напомним, что вязкость обратно пропорционально зависит от ионной электропроводности. Разработка химических источников тока, обладающих пониженной вязкостью, автоматически повысит их электропроводность. Ионная электропроводность является одним из важнейших дескриптором электролита, а ее высокое значение делает рассматриваемую ион-молекулярную систему технологически конкурентной [28].

Разработка новых электролитных растворов с улучшенными свойствами, такими как высокая ионная проводимость, широкое электрохимическое окно, низкая вязкость, высокая термическая и химическая стабильность, является критически важной задачей для создания более эффективных, безопасных и долговечных электрохимических устройств [28-29]. Для достижения этой цели необходимо глубокое понимание взаимосвязи между составом, структурой и свойствами электролитов на молекулярном уровне организации материи.

Традиционные экспериментальные методы, такие как измерение проводимости, вязкости, плотности и электрохимической стабильности, предоставляют ценную информацию о макроскопических свойствах электролитов. Однако они не всегда позволяют получить полное представление о молекулярных механизмах, лежащих в основе наблюдаемых явлений. В последние годы исследование ландшафта потенциальной энергии (ЛПЭ) стало мощным инструментом для изучения электролитных растворов и рационального улучшения их характеристик. Анализ ЛПЭ, основанный на применении методов молекулярного моделирования, таких как квантово-химические расчеты и классическая молекулярная динамика, позволяет получить детальную информацию о взаимодействиях между компонентами раствора, т.е. ион-молекулярных и межмолекулярных электростатических силах. Последние и определяют равновесные свойства электролитного раствора [30-33].

Катионы, анионы, молекулы растворителя и молекулярные добавки влияют на макроскопические свойства электролита. Например, исследование ЛПЭ может выявить предпочтительные конфигурации сольватов ионов, энергетические барьеры для активации ионного транспорта, а также влияние температуры и состава на структуру, транспорт и макроскопические свойства электролита [34-36].

В данной статье нами представлен критический обзор недавних примеров применения методов исследования ЛПЭ, основанных на классических и квантово-химических алгоритмах, для анализа структуры, динамики и термодинамических свойств электролитных растворов [5,22,37-38].

## Обсуждение изученной литературы

Теоретические расчеты в теоретической химии и молекулярной физике являются мощным инструментом для исследования ЛПЭ электролитных растворов. Они позволяют моделировать взаимодействия между компонентами системы на атомарном уровне и получать детальную термодинамическую информацию о структуре и динамике раствора [39-42]. Впоследствии термодинамические дескрипторы ЛПЭ могут быть переведены на язык электрохимических и физико-химических свойств, связанных с экспериментально полученными свойствами [5].

Численные методы расчета потенциальной энергии ион-электронной системы из первых принципов, основанные на фундаментальных законах квантовой механики, позволяют получать высокоточные результаты для параметризации ЛПЭ. К сожалению, они требуют значительных вычислительных ресурсов и не могут быть применены к системам значительных размеров вследствие принципиальной невозможности получить аналитическое решение волнового уравнения для многоэлектронных задач. Метод функционала плотности (DFT) представляет собой менее ресурсоемкий метод, который широко используется для расчета электронной структуры и свойств молекул, термодинамических фаз и материалов [7,43-45]. Результатом расчета электролитной системы методом DFT является как полная потенциальная энергия, так и силы, действующие на каждый атом, катион или анион.

Классическая молекулярная динамика (МД) использует классические законы движения для моделирования движения атомов и молекул [46-47]. Классическая МД не рассматривает электроны в качестве отдельных центров взаимодействия. Атомы взаимодействуют между собой посредством простых потенциалов [48]. Параметризация потенциальных функций производится заранее на основании квантово-химических или экспериментальных данных. В процессе МД моделирования потенциальная энергия системы зависит только от расстояния между каждыми двумя взаимодействующими центрами в текущий момент времени. Этот метод позволяет исследовать динамику системы и получать информацию о ее транспортных свойствах. Например, в рамках классического МД моделирования существуют методики для расчета самодиффузии молекул растворителя и проводящей подсистемы электролита, ионной проводимости и вязкости многокомпонентной электролитной системы как целого [46-47,49-50].

Ab initio МД объединяет принципы квантовой механики и молекулярной динамики, что позволяет моделировать химические реакции и другие процессы, связанные с изменением электронной структуры. Важным практическим преимуществом Ab initio МД является способность корректно описывать эффекты электронной поляризации и частичного смещения электронной плотности без предварительной трудоемкой параметризации этого физического процесса для разных сочетаний функциональных групп [10,51]. Электронная поляризация представляет собой крайне распространенное явление в ион-молекулярных системах, способное существенно влиять на структуры сольватных комплексов и термодинамику самого процесса сольватации. Величина электронной поляризации прямо пропорциональна отличию формального заряда участвующего в ней катиона или аниона от нуля. Тогда как метод Ab initio МД способен давать более реалистичные результаты моделирования, его стоимость на порядки превышает стоимость классического МД моделирования. Потому метод Ab initio МД может быть использован лишь для отдельно выделенных ионных кластеров и ионных сольватов [10].

Метод Монте-Карло Метрополиса является альтернативным молекулярной динамике способом движения системы по ЛПЭ [52-55]. Данный метод используется для генерации конфигураций системы, соответствующих заданному распределению вероятностей. Система следует в том направлении, принадлежащем ЛПЭ, которому соответствует уменьшение потенциальной энергии системы [56]. В отличие от МД, эволюция системы не требует вычисления градиентов энергии (действующих сна атомы сил) и не использует временной шаг интегрирования уравнений движения. Преимущественной областью применения метода оказались равновесные физико-химические системы [39,53,56-59]. Следовательно, наиболее естественным применением метода Монте-Карло Метрополиса видится возможность исследовать термодинамические свойства и фазовые переходы [6,60-61].

Метод инжекции кинетической энергии позволяет целенаправленно искать низкоэнергетические стационарные точки путем периодического кинетического возбуждения исследуемой системы [12,62-63]. Данный метод комбинирует равновесную полуэмпирическую МД со стохастическим компонентом. Тогда как МД моделирование поступательно ведет систему к ее минимальным потенциальным энергиям, искусственно вносимое возбуждение позволяет не задерживаться в локальных потенциальных долинах. Таким образом, система довольно быстро оказывается в реалистичных микроскопических ион-молекулярных конфигурациях вне зависимости от энергии произвольно выбираемого начального состояния. Стоит отметить, что определенная модификация импульсов частиц не приводит к образованию нефизичных ион-молекулярных конфигураций, но в то же время изменяет текущую фазовую траекторию системы [10,64-66]. Таким образом, инжекция кинетической энергии по сравнению со стандартным методом МД позволяет системе посетить большее количество микроскопических состояний собственного фазового пространства (рис. 2).

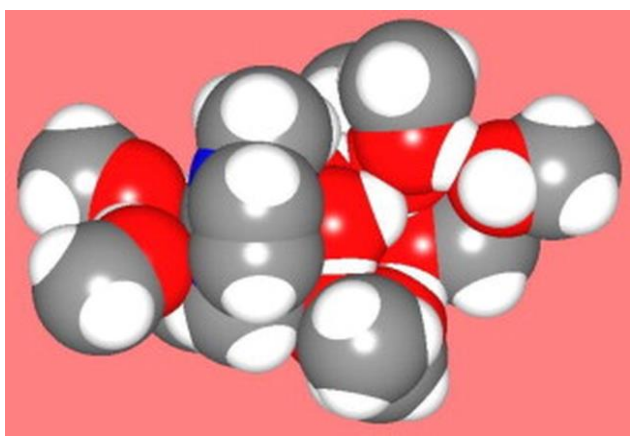

Рис. 2. Сольватная оболочка катиона имидазолия в смеси воды и метанола. Воспроизведено из авторского источника [67] с разрешения Элзевир. Copyright Elsevier (2024).

Экспериментальные методы исследования предоставляют важную информацию о структуре и динамике электролитных растворов, которая может быть использована для валидации и уточнения теоретических моделей. Рентгеновская дифракция позволяет определить пространственное расположение атомов и молекул в растворе [68]. Нейтронная дифракция чувствительна к легким атомам, таким как водород, что делает ее полезной для исследования структуры растворителей [38]. Спектроскопия ядерного магнитного резонанса предоставляет информацию о локальной структуре и динамике молекул в растворе [69]. Рамановская спектроскопия используется для изучения колебательных мод молекул и получения информации о межмолекулярных взаимодействиях. Электрохимические методы, такие как циклическая вольтамперометрия, импедансная спектроскопия и прочие, позволяют исследовать электрохимические свойства растворов и кинетику электродных реакций [66].

ЛПЭ оказывает принципиальное влияние на различные свойства электролитных растворов, которые определяют их эффективность в различных приложениях. Ионная проводимость является одним из ключевых параметров электролитов, определяющим эффективность переноса заряда в электрохимических устройствах. ЛПЭ влияет на ионную проводимость через следующим образом. ЛПЭ определяет энергетические барьеры, которые ионы должны преодолеть для перемещения в растворе. Чем ниже энергия активации, тем выше ионная подвижность, а, следовательно, и проводимость электролитного раствора. ЛПЭ отражает координацию ионов растворителем, что в свою очередь влияет на их подвижность. ЛПЭ содержит информацию о силе взаимодействия между ионами, образовании ионных пар. Спаривание катионов и анионов в электролитном растворе видится нежелательным в контексте практического применения, поскольку приводит к снижению ионной проводимости. Дизайн электролита включает подбор оптимальных концентраций всех его компонентов с целью минимизации образования ионных агрегатов. Зависимость ионной электропроводности от мольной доли ионов в электролите содержит максимум, соответствующих желательным свойствам системы.

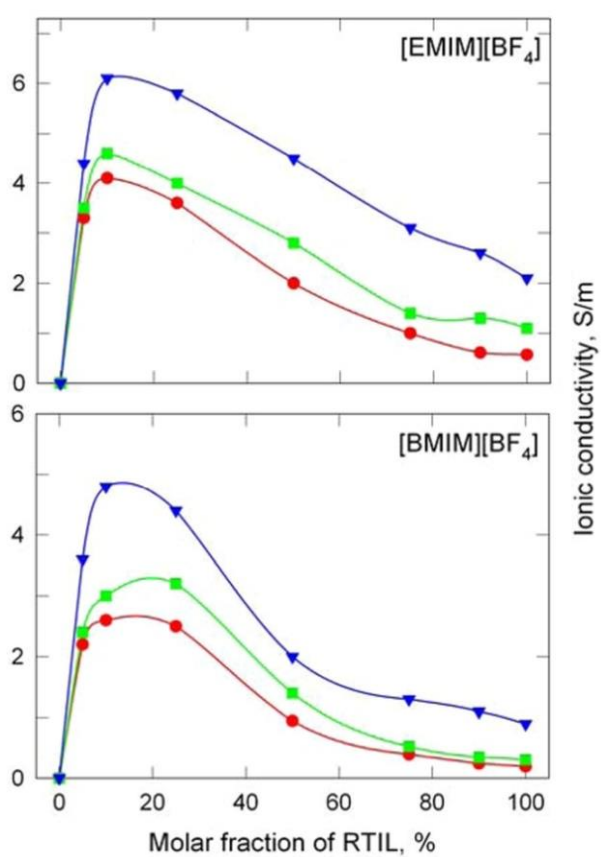

Рис. 3. Рассчитанный максимум электроводности в смесях ионных жидкостей, [EMIM][BF4] и [BMIM][BF4], с ацетонитрилом при 283 K (красные кружки), 298 K (зеленые квадраты) и 323 K (синие треугольники). Воспроизведено из авторского источника [70] с разрешения Американского химического общества. Copyright ACS (2024).

Электрохимическая стабильность электролита определяет его способность противостоять разложению при высоких напряжениях, что критически важно для работы аккумуляторов и других устройств [71]. Форма некоторых участков ЛПЭ и конкретные разницы значений энергии системы описывает электрохимическую стабильность. В частности, ЛПЭ определяет энергию, необходимую для разложения молекул растворителя или других компонентов электролита. Взаимодействие с электродами также может быть отражено в ЛПЭ, если таковые были включены в модельную систему в процессе теоретического исследования ландшафта. ЛПЭ определяет энергию взаимодействие электролита с электродами и дает оценку возможности образования пассивирующих слоев или протекания других нежелательных реакций [72].

Вязкость электролита влияет на транспортные свойства и эффективность работы электрохимических устройств [73-74]. Вязкость связана с градиентами энергии, наблюдаемыми на ЛПЭ. Межмолекулярные взаимодействия, определяющие ЛПЭ, характеризуют силу взаимодействия между молекулами растворителя и ионами, что влияет на вязкость раствора.

ЛПЭ отражает возможные микроскопические состояния электролитного раствора при разные температурах и давлениях. Структура электролитного раствора, среди прочего, подразумевает образование кластеров или других упорядоченных структур, что также может кардинальным образом влиять на вязкость.

Таким образом, информация о ЛПЭ может быть использована для рационального дизайна новых электролитных растворов с улучшенными характеристиками. Анализ ЛПЭ позволяет выбрать наиболее подходящий растворитель, который обеспечивает оптимальную сольватацию ионов и минимальную энергию активации для ионного транспорта. ЛПЭ позволяет количественно оценить силу взаимодействия между ионами и выбрать соль, которая минимизирует ассоциацию ионов и обеспечивает высокую ионную проводимость. ЛПЭ позволяет исследовать влияние добавок на структуру и свойства электролита, например, на вязкость, электрохимическую стабильность и смачиваемость электродов [75].

На протяжении последних двух десятилетий ионные жидкости рассматриваются в качестве перспективной основы для жидких электролитных систем. Низкая летучесть этих соединений исключает потерю электролита на любом этапе производства и эксплуатации химического источника тока. Некоторые ионные жидкости обладают лишь незначительной токсичностью для человека и животных или вообще признаны безвредными. Широкое диапазон жидкого состояния, особенно при низких температурах, отрывает новые рубежи применений таких электролитов. В силу того, что ионные жидкости не содержат молекул, использования их в качестве электролитов позволяет достичь доселе небывалых концентрации катионов и анионов в системе. Тем не менее текущей проблемой является высокая вязкость и, соответственно, посредственная электропроводность данных химических соединений. Для стимулирования ионного транспорта в электролитах разрабатываются оптимальные составы смесей ионных и молекулярных жидкостей, например, включающие ацетонитрил, метанол или пропиленкарбонат [76]. ЛПЭ позволяет исследовать структуру и свойства ионных жидкостей, равно как и вышеописанных смесей, которые остаются перспективными электролитами для различных электрохимических приложений [70].

Выходя за рамки тематики электролитных растворов, следует упомянуть об универсальности ЛПЭ как источника информации о широком спектре химических и физико-химических свойств прототипированной системы. Аналогично описанным выше случаям, исследованная ЛПЭ позволяет исследовать механизмы ионного транспорта в твердых и полимерных электролитах и разрабатывать новые материалы с более высокой ионной проводимостью. Этой цели в перспективе можно достичь путем подавления нежелательных атом-атомных взаимодействий. Разумеется, такая модификация электролитной системы требует также и синтетических усилий со стороны электрохимической лаборатории.

Миграция ионов в жидких электролитных растворах представляет собой сложный физический процесс, который требует преодоления определенных энергетических барьеров. Эти барьеры возникают из-за взаимодействия мигрирующего иона с окружающими атомами или молекулами. Для миграции ионов в электролите необходимо преодолеть несколько значимых энергетических барьеров различной высоты. Конкретные значения активационных барьеров в рамках ионного транспорта зависит от природы растворителя и электростатического заряда иона. Например, хорошо известно, что сольватация катионов в абсолютном большинстве случаев описывается более значительными потенциальными энергиями, чем сольватация анионов [77].

С помощью ЛПЭ возможно рассчитать энергию активации для выхода иона из сольватной оболочки. Ионы в электролите окружены молекулами растворителя, образуя сольватную оболочку. Чтобы ион начал двигаться, ему нужно разорвать эти связи, что требует энергии. Далее необходимо преодолеть энергетический барьер для движения иона через электролит. Движение иона в электролите затруднено из-за взаимодействия с другими ионами и молекулами растворителя. Ион должен преодолеть силы электростатического притяжения и отталкивания, а также силы Ван-дер-Ваальса. Энергетический барьер сопутствует так же и входу иона в его новую сольватную оболочку. Когда ион достигает места назначения, ему нужно образовать новые связи с молекулами растворителя, чтобы создать новую сольватную оболочку. Корректно рассчитанная ЛПЭ содержит полную термодинамическую информацию, характеризующие описанные выше процессы [78].

На высоты энергетических барьеров влияют следующие факторы. Тип растворителя и растворенных веществ изначально детерминирует энергетические барьеры. Например, вязкость растворителя влияет на энергию, необходимую для движения иона. Чем выше концентрация ионов, тем сильнее взаимодействие между ними, что увеличивает энергетические барьеры. Повышение температуры увеличивает кинетическую энергию ионов, что помогает им преодолевать энергетические барьеры. Внешнее электрическое поле может снижать энергетические барьеры для движения ионов в определенном направлении.

Величина энергетических барьеров в ЛПЭ на микроуровне определяет макроскопические характеристики электролита. Чем ниже энергетические барьеры, тем выше подвижность ионов и, следовательно, ионная проводимость электролита. Энергетические барьеры влияют на скорость диффузии ионов в электролите, динамику зарядки и разрядки [78]. Понимание энергетических барьеров для миграции ионов ключевые для разработки новых электролитов с улучшенными характеристиками, например, для использования в батареях, топливных элементах и других электрохимических устройствах.

ЛПЭ позволяет моделировать процессы, происходящие в аккумуляторах [79], такие как интеркаляция ионов, образование пассивирующих слоев и деградация электролита. Более того, ЛПЭ позволяет моделировать процессы переноса заряда и массы в топливных элементах и оптимизировать их конструкцию. Наконец, ЛПЭ позволяет моделировать процессы накопления заряда в суперконденсаторах и разрабатывать новые материалы с высокой емкостью [78].

Исследование ЛПЭ позволило разработать новые электролиты с высокой ионной проводимостью и электрохимической стабильностью, что привело к созданию аккумуляторов с повышенной эффективностью, емкостью, сниженной ценой производства и увеличенным сроком службы. Аналогичные исследования суперконденсаторов позволили повысить скорости их емкости и темпы зарядки [80-81].

Литий-серные аккумуляторы обладают высокой теоретической емкостью, но их практическое применение ограничено рядом проблем, таких как растворение полисульфидов лития и низкая электропроводность серы. Исследование ЛПЭ позволяет понять механизмы этих процессов и разработать электролиты, которые минимизируют растворение полисульфидов и повышают электропроводность серы. Например, использование электролитов с высокой долей соли или добавлением специальных присадок может существенно улучшить характеристики литий-серных аккумуляторов [82].

Натрий-ионные аккумуляторы рассматриваются как перспективная альтернатива литий-ионным аккумуляторам, благодаря более низкой стоимости и большей доступности натрия. Однако, ионы натрия имеют больший размер и массу по сравнению с ионами лития, что приводит к снижению ионной проводимости и ухудшению кинетики электродных реакций. Исследование ЛПЭ позволяет оптимизировать состав электролитов для натрий-ионных аккумуляторов, например, путем выбора растворителей с высокой диэлектрической проницаемостью и низкой вязкостью, а также добавления солей с высокой растворимостью и низкой энергией активации для транспорта ионов натрия [10,51]. Моделирование ЛПЭ позволяет включить в рассмотрение электрод и энергетически описать поведение электролита в состоянии заряженного устройства (рис. 4).

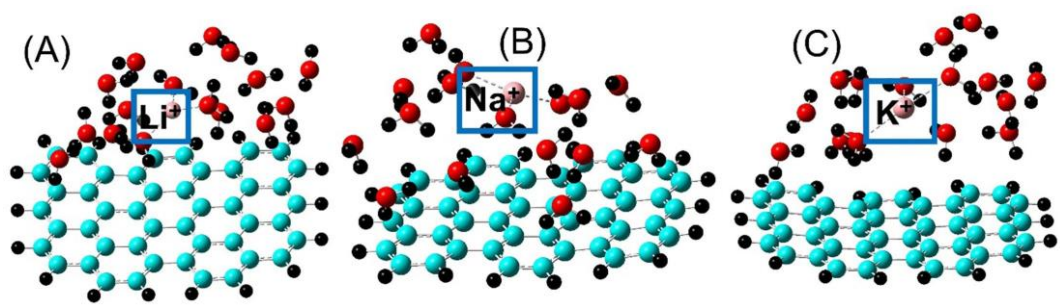

Рис. 4. Низкоэнергетические ион-молекулярные конфигурации инов лития, натрия и калия на поверхности отрицательно заряженного графенового электрода. Пунктирные линии показывают ближайшие к ионам молекулы воды. Воспроизведено из авторского источника [51] с разрешения Элзевир. Copyright Elsevier (2024).

Исследование влияния температуры и давления на ЛПЭ позволяет понять, как изменяются свойства электролитов в различных условиях эксплуатации. Учет формы ЛПЭ в наложенном электрическом поле позволяет понять механизмы переноса заряда в электролитах и оптимизировать работу электрохимических устройств [83]. Взаимодействия электролита с поверхностью электрода важны для обнаружения механизмов электродных реакций. Таким образом, представляется возможным предложить доработанные новые материалы с улучшенными электрохимическими свойствами.

Развитие теоретических методов исследования ЛПЭ открывает новые возможности для разработки электролитных растворов с улучшенными характеристиками [84-85]. Прогресс в методах расчета потенциальной энергии системы может позволить изучать электролитные системы более сложной композиции и получать больше информации о ЛПЭ. Сочетание теоретических расчетов и экспериментальных данных позволит получить более симбиотическое представление о ЛПЭ и использовать эту информацию для рационального дизайна новых электролитов. Наконец, применение методов машинного обучения для анализа ЛПЭ позволит ускорить процесс разработки новых электролитов и оптимизировать их свойства [83].

## Выводы

В настоящей работе нами проведен анализ значительного количества наиболее современных литературных источников, касающихся как полноценного исследования электролитных растворов методами молекулярного моделирования для описания ЛПЭ, так и вспомогательных компьютерных расчетов для подтверждения того или иного экспериментально полученного результата.

На основании проведенного критического анализа полагаем корректным сделать следующие выводы. Исследование ландшафта потенциальной энергии представляет собой мощный инструмент для анализа и прогнозирования свойств электролитных растворов. Применение методов исследования, основанных на классических и квантово-химических подходах, позволяет получить уникальную физическую информацию. Во-первых, видится возможным детально изучать межмолекулярные и ион-молекулярные взаимодействия, определяющие макроскопические свойства электролитного раствора. Ионная электропроводность, вязкость и плотность могут быть рассчитаны с помощью аналитических соотношений на основании статистической обработки фазовой траектории системы. Во-вторых, несложно идентифицировать стабильные конфигурации ионов и сольватов, влияющие на равновесное состояние электролита и механизмы ионного транспорта в последней. В-третьих, получаемые термодинамические свойства моделируемой системы и энергий парных ион-молекулярных взаимодействий позволяют оценить барьеры для миграции ионов, что напрямую связано с проводимостью электролита.

Таким образом, сочетание методов исследования ЛПЭ с экспериментальными данными открывает новые перспективы для целенаправленного дизайна электролитных растворов с заданными характеристиками, что крайне важно для разработки высокоэффективных электрохимических устройств.

## Конфликт интересов

Автор заявляет об отсутствии финансовых и профессиональных интересов или связей, которые бы могли повлиять на анализ интерпретируемых результатов.

## СПИСОК ИСТОЧНИКОВ

[1] **Zittlau, P.; Mross, S.; Gond, D.; Kohns, M.**, Molecular modeling and simulation of organic electrolyte solutions for lithium ion batteries // J Chem Phys, 161 (12) (2024) 124118. 10.1063/5.0228158.

[2] **Shi, N.; Wang, G.; Mu, T.; Li, H.; Liu, R.; Yang, J.**, Long side-chain imidazolium functionalized poly(vinyl chloride) membranes with low cost and high performance for vanadium redox flow batteries // J Mol Liq, 376 (2023) 121401. 10.1016/j.molliq.2023.121401.

[3] **Deb, D.; Bhattacharya, S.**, Imidazolium-based ionanofluid electrolytes with viscosity decoupled ion transport properties for lithium-ion batteries // J Mol Liq, 379 (2023) 121645. 10.1016/j.molliq.2023.121645.

[4] **Wu, Z.; Tian, Y.; Chen, H.; Wang, L.; Qian, S.; Wu, T.; Zhang, S.; Lu, J.**, Evolving aprotic Li-air batteries // Chem Soc Rev, 51 (18) (2022) 8045-8101. 10.1039/d2cs00003b.

[5] **Habibi, P.; Polat, H. M.; Blazquez, S.; Vega, C.; Dey, P.; Vlugt, T. J. H.; Moultos, O. A.**, Accurate Free Energies of Aqueous Electrolyte Solutions from Molecular Simulations with Non-polarizable Force Fields // J Phys Chem Lett, 15 (16) (2024) 4477-4485. 10.1021/acs.jpclett.4c00428.

[6] **Lu, G. W.; Li, C. X.; Wang, W. C.; Wang, Z. H.**, A Monte Carlo simulation on structure and thermodynamics of potassium nitrate electrolyte solution // Mol Phys, 103 (5) (2005) 599-610. 10.1080/00268970410001683834.

[7] **Parida, R.; Yong Lee, J.**, Boron based podand molecule as an anion receptor additive in Li-ion battery electrolytes: A combined density functional theory and molecular dynamics study // J Mol Liq, 384 (2023) 122236. 10.1016/j.molliq.2023.122236.

[8] **Kartha, T. R.; Mallik, B. S.**, Revisiting LiClO4 as an electrolyte for Li-ion battery: Effect of aggregation behavior on ion-pairing dynamics and conductance // J Mol Liq, 302 (2020) 112536. 10.1016/j.molliq.2020.112536.

[9] **Jiang, H.; Zhang, Q.; Zhang, Y.; Sui, L.; Wu, G.; Yuan, K.; Yang, X.**, Li-Ion solvation in propylene carbonate electrolytes determined by molecular rotational measurements // Phys Chem Chem Phys, 21 (20) (2019) 10417-10422. 10.1039/c8cp07552b.

[10] **Chaban, V. V.; Andreeva, N. A.**, Shorter-chained trialkylsulfonium cations are preferable as admixtures to lithium-ion and sodium-ion electrolytes in acetonitrile // J Mol Liq, 385 (2023) 122399. 10.1016/j.molliq.2023.122399.
20